\documentclass{article}
\usepackage{graphicx}
\usepackage[final]{corl_2023} 

\usepackage{floatrow}
\newfloatcommand{capbtabbox}{table}[][\FBwidth]
\usepackage{times}
\usepackage{tabularx}
\usepackage{float}
\usepackage{paralist}
\usepackage[inline]{enumitem}
\usepackage{comment}
\usepackage{soul}
\usepackage{url}
\usepackage{color}

\usepackage[utf8]{inputenc}
\usepackage[small]{caption}
\usepackage{graphicx}
\usepackage{amsmath}
\usepackage{amsthm}
\usepackage{booktabs}
\usepackage{algorithm}
\usepackage{algorithmic}
\usepackage{multirow}
\usepackage{subcaption}
\usepackage{amsfonts}
\usepackage{sidecap}

\title{Multi-Agent Training for Pommerman: Curriculum Learning and Population-based Self-Play Approach}

\author{
  Nhat-Minh Huynh$^{1,*}$ \quad    Hoang-Giang Cao$^{2,3,*}$ \quad  I-Chen Wu$^{4,\dagger}$
  \\
  $^1$Agoda (Booking Holdings Inc.), Thailand \\
  $^2$Ming Chi University of Technology, Taiwan \\
  $^3$Can Tho University, Vietnam \\
  $^4$National Yang Ming Chiao Tung University, Taiwan
}

\begin{document}
\maketitle
\def\thefootnote{$^{*}$}\footnotetext{Equal contribution. 
  \quad $^{\dagger}$Corresponding.}\def\thefootnote{\arabic{footnote}}



\begin{abstract}
Pommerman is a multi-agent environment that has received considerable attention from researchers in recent years.
This environment is an ideal benchmark for multi-agent training, providing a battleground for two teams with communication capabilities among allied agents.
Pommerman presents significant challenges for model-free reinforcement learning due to delayed action effects, sparse rewards, and false positives, where opponent players can lose due to their own mistakes.
This study introduces a system designed to train multi-agent systems to play Pommerman using a combination of curriculum learning and population-based self-play.
We also tackle two challenging problems when deploying the multi-agent training system for competitive games: sparse reward and suitable matchmaking mechanism.
Specifically, we propose an adaptive annealing factor based on agents’ performance to adjust the dense exploration reward during training dynamically.
Additionally, we implement a matchmaking mechanism utilizing the Elo rating system to pair agents effectively.
Our experimental results demonstrate that our trained agent can outperform top learning agents without requiring communication among allied agents.
\end{abstract}


\section{Introduction}
\label{sec:introduction}
In 1983, Bomberman was released for the Nintendo Entertainment System (NES) platform.
In 2018, Pommerman \cite{resnick2018pommerman} was introduced; it is a well-organized open-source environment that is implemented in Python.
This version enhances the original Bomberman with a battle scenario where players are allowed to communicate in a team.

Pommerman is an appropriate multi-agent benchmark because the environment offers various challenges, including multiple agent interaction, sparse rewards, false-positive rewards, delayed action effects, and complex exploration \cite{ChaoCao-onhard}.
There are four agents in a game, either playing in 2vs2 team mode or individually competing against each other in Free-For-All mode.
The sparse reward is a problem related to the time that an agent receives a reward.
In Pommerman, a game reward of win or loss is given to each agent at the end of an episode, which can take up to 800 timesteps.
The false-positive reward occurs when rewards are not derived from appropriate actions by an agent but from the opponent's mistake of committing suicide.
Another challenging problem in Pommerman is the delay action effect.
In the game, the only way to explore the board and eliminate an enemy is by placing bombs, but the effect of bomb placement does not produce a direct result but delays 10 timesteps.
In addition, as the vision of an agent is limited to a 9x9 grid around it, the environment turns into a partially observable environment, which makes it hard for an agent to explore the board or find enemies.



\begin{figure}[!ht]
\includegraphics[width=0.8\textwidth]{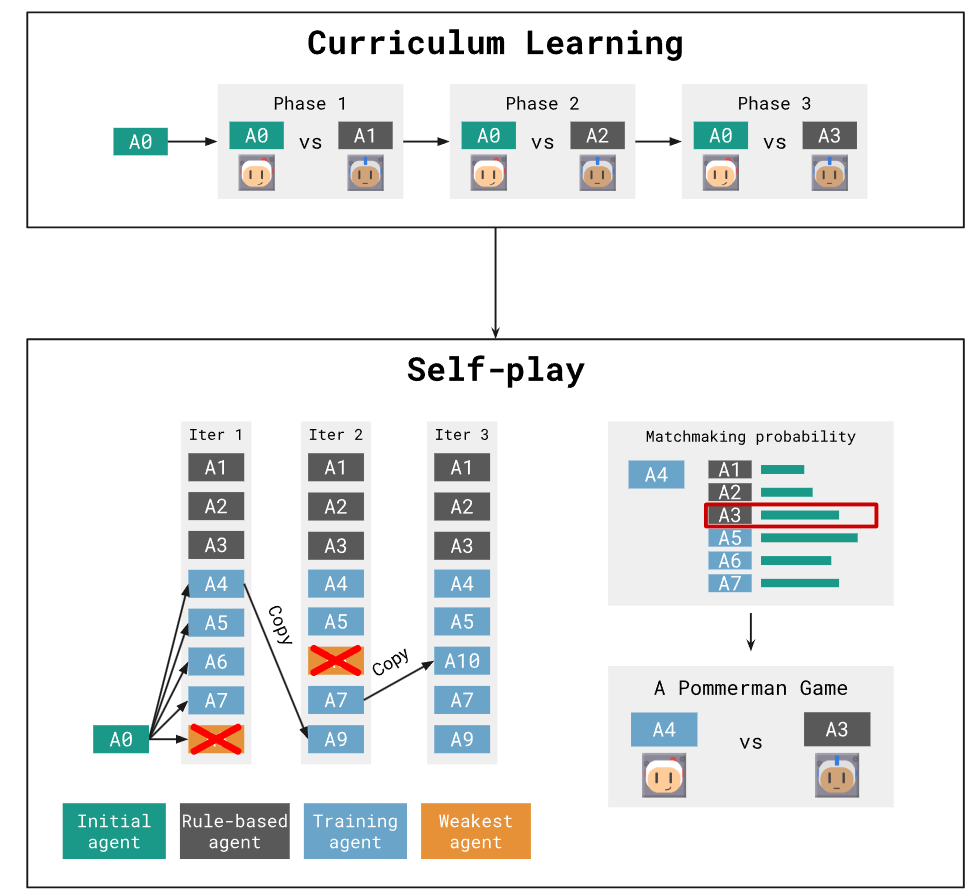}
\centering
\caption{Overview of our multi-agent training system with two stages: curriculum learning and population-based self-play.}
\label{fig:overview_system}
\end{figure}

In this study, we introduce a multi-agent training system to play the 2vs2 team mode of Pommerman.
Our system includes two stages: curriculum learning and population-based self-play.
The curriculum learning, consisting of three incremental difficulty phases, assists the agent in learning essential skills to handle the games, such as exploring the map, picking up items, hiding from an explosion, and using defensive strategies to stay safe and survive while simultaneously fighting to eliminate enemies.
Once agents acquire those necessary skills, we design a population-based self-play system wherein a population of agents compete against each other, naturally evolving their own strategies to improve performance.
Figure \ref{fig:overview_system} shows the overview of our system.

Deploying a multi-agent self-play training system for a competitive game poses two challenges \cite{Trapit_Emergent}: 
(1) addressing exploration problems in a competitive game with sparse reward; and
(2) designing a suitable matchmaking mechanism for effective agent pairing in training, enabling progressive learning.
To address the exploration problem with sparse reward, we introduce an adaptive annealing factor, which dynamically anneals the dense exploration reward during training based on the agent's performance.
To design a suitable matchmaking, we implement the matchmaking probability based on the Elo rating system, ensuring incremental learning.

\noindent The main contributions of this paper are summarized as follows:
\begin{inparaenum}[1)]
    \item We present a multi-agent training system to play Pommerman, which includes two stages: curriculum learning and population-based self-play. 
    \item  We propose an adaptive annealing factor based on agents' performance to dynamically adjust the dense exploration reward during training.
    \item  We implement a matchmaking mechanism utilizing the Elo rating system to pair agents effectively
     \item In our experiments, we demonstrate that our trained agent can outperform learning agents, even without requiring communication among allied agents.
\end{inparaenum}

\section{Approaches}
 

\subsection{Curriculum Learning Stage}
\label{sec:approach_curriculum}
\subsubsection{Curriculum Learning with Incremental Difficulty Agents}
Pommerman is a complex environment that requires multiple skills to handle the game.
To encourage the agent to acquire different skills, we designed a curriculum learning with three phases: exploring the map and locating opponents, eliminating opponents, and surviving encounters while fighting opponents.
Three phases correspond to three rule-based agents (inspired by Chao Gao et al. \cite{ChaoCao-Skynet}), denoted as \texttt{static\_agent}, \texttt{simple\_moving\_agent} and \texttt{simple\_bomb\_agent}.

In the first phase, the training agent plays with the \texttt{static\_agent}, who idle and are waiting to be eliminated by our agent.
In this phase, the agent learns how to explode wooden walls to open up the passages, pick up items, hide from an explosion of its own bombs, and finally, eliminate the static opponents.

In the second phase, the training agent plays with \texttt{simple\_moving\_agent} who move randomly on the board but are not allowed to place bombs.
Furthermore, the opponents are programmed to dodge the bomb explosion safely.
As the difficulty increases, our agent is required to find strategies to place bombs more effectively.

In the third phase, the \texttt{simple\_bomb\_agent} are able to move and place bombs randomly.
In this scenario, approaching the opponents becomes more challenging for our agent.
Therefore, our agent is required to learn defensive strategies to stay safe and survive while simultaneously fighting to eliminate enemies.

The next phase of the curriculum learning stage is activated when the training agent achieves a 55\% win rate over the current rule-based agent.

\subsubsection{Adaptive Exploration Reward by Performance}
The main objective of Pommerman is to eliminate the opponents.
However, players are initially located in the corners of the map and are blocked by wooden walls and stones.
Thus, the gameplay can be divided into two subtasks:
(1) explore the map to navigate to the opponent; and
(2) strategically place bombs to engage in fighting and eliminate them.
These two sub-tasks correspond to two types of reward: exploration reward and game reward, denoted as $e$ and $R$, respectively.
The details of reward setting are detailed in \ref{sec:reward_system}.

There is a conflict between using exploration rewards and game rewards. 
The dense exploration reward is necessary at the beginning of training to encourage the agent to explore the map and open up the passages.
However, using dense exploration rewards distracts the purpose of the self-evolving strategy, which benefits from the sparse game reward.
Conversely, directly using the sparse game reward is excessively challenging for agents to learn gameplay effectively due to the complexity of exploration in the game environment.

To solve this problem, a work \cite{Trapit_Emergent} calculated the reward function with an annealing factor, denoted as $\alpha$, to linearly reduce the dense exploration reward to zero during the training. In the work \cite{Trapit_Emergent}, the reward function at timestep t is defined as follows:
\begin{equation}
    r_t = \alpha_t e_t + (1-\alpha_t)R
    \label{equal:adaptive_reward}
\end{equation}
where $r_t$ is the total reward agent received, $e_t$ is the exploration reward, $R$ is a game reward, and $\alpha_t$ is the annealing factor, controlling the impact of $e_t$ and  $R$.
Generally, deciding when the annealing factor should equal zero is a manual fine-tuning process.

\begin{figure}[!ht]
\includegraphics[width=0.5\textwidth]{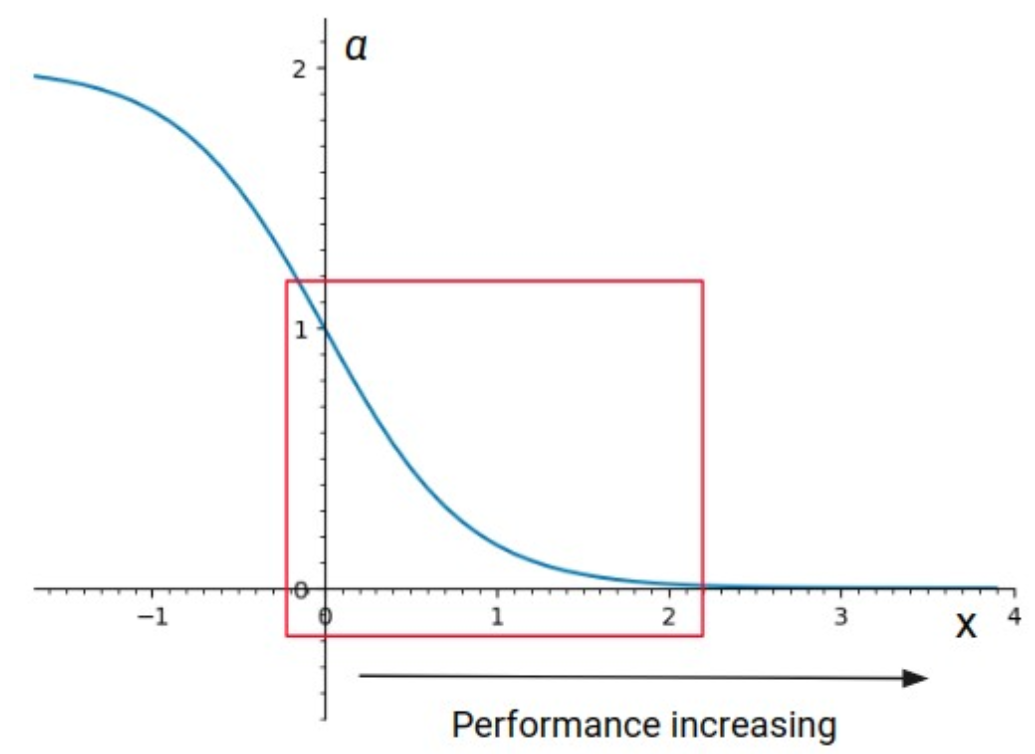}
\centering
\caption{Annealing factor $\alpha$ function with k = 1.2.
Noted annealing factor only calculated in the range of [0, 2], which is the part in red square}
\label{fig:tanh_alpha}
\end{figure}

To dynamically adjust the annealing factor $\alpha$ for adapting to strategy changes during the training, we propose an adaptive annealing factor based on the agent's performance as follows:
\begin{equation}
    \alpha = 1-tanh (k*x)
\end{equation}
where $k$ is the tuning parameter, and $x$ is the current performance of an agent.
In our work, an agent's performance is measured by the average number of deaths of the enemy so that $x$ is in the range of [0, 2] and $k$ is set to 1.2 to obtain a suitable curve of $tanh$ function (shown in Figure \ref{fig:tanh_alpha}). 

At the beginning of training, the agent is not able to approach and eliminate the opponent; the current performance of an agent  $x = 0$. 
Thus, the exploration reward fully impacts the reward, motivating the agent to place bombs to explore the map.
Subsequently, the agent eventually locates the opponents to eliminate them, increasing the performance.
As performance increases, the impact of the dense exploration rewards and sparse game rewards shifts.
This encourages the agent to gradually prioritize self-evolving strategies aligned with the primary game objective.

After the curriculum learning stage, the dense exploration reward annealed to zero, entirely replaced by the sparse game reward.
So, in the self-play stage, the agent is trained only on game rewards.

\subsection{Population-based Self-play Stage}
\label{sec:approach_self_play}
\subsubsection{Population-based Self-play System}
After training in the curriculum learning stage, the agent knows the properties of the Pommerman environment and learns the essential skills to handle the game.
To improve the agent's strength, we designed a population-based self-play system where a population of agents plays against each other for self-improvement.

We designed a population-based self-play system with eight agents.
Three agents are rule-based agents used in the curriculum learning stage, to prevent from forgetting learned skills
The remaining five learning-based agents are initialized by the trained agent from the curriculum learning stage.

During the self-play training, if an agent has a win rate below 45\% over the population, it is considered to be replaced by a stronger agent.
A stronger agent is selected randomly from the four remaining agents.
Figure \ref{fig:overview_system} shows the design of our population-based self-play stage.

To ensure effective progressive learning during self-play, selecting an appropriate opponent for pairing is crucial.
In the next subsection, we will introduce a matchmaking mechanism based on Elo rating to tackle this problem.

\subsubsection{Match Making Probability}
\label{sec:matchmaking}
During the self-play stage, each agent gains experience by interacting with other agents.
Therefore, implementing a suitable matchmaking system is necessary to effectively pair agents for training, ensuring they facilitate progressive learning, adapt to new strategies introduced by other agents, and avoid getting stuck \cite{Trapit_Emergent}.

In this research, we implement the matchmaking probability based on the Elo rating system.
The Elo rating system is commonly used in board games like Chess and Go \cite{elo1978rating}.
In competitive games, the Elo rating is adopted to measure players' strength and calculate the win rate of a match between two players.


Let the Elo ratings of agent A and agent B be denoted as $R_A$ and $R_B$, respectively.
The formula to calculate the expected win rate of Agent A against Agent B is defined as follows:
\begin{equation}
    E_A = \frac{1}{1+10^{(R_B-R_A)/400}}
    \label{eqal:expected_score}
\end{equation}

Then, after a match between agent A and agent B, the Elo rating of agent A is updated as follows:
\begin{equation}
    R'_A = R_A + K(S_A - E_A)
    \label{eqal:elo_formula}
\end{equation}
where $K$ is the maximum adjustment per game, and $S_A$ is the actual score of agent A after a match: 1, 0.5, or 0 if the match results in a win, tie, or loss, respectively.

In a multi-agent training system, the matchmaking probability of other agents to a selected agent is calculated by applying the softmax function to the expected win rate between those agents and the selected one.
As a result, a higher Elo rating agent has a higher chance of being selected as an opponent of the current training agent.

For example, a population-based self-play system with 4 agents, A1, A2, A3, and A4, with 1010, 1020, 920, and 986 Elo points corresponding.
If agent A4 is the current training agent to play a match, the expected win rate of A1, A2, and A3 are 0.53, 0.54, and 0.4 (Equation \ref{eqal:expected_score}).
We then apply the soft-max function to these expected win rates to obtain the matchmaking probabilities of agents A1, A2, and A3 as follows: 0.346, 0.35, and 0.304.


\section{Experiment Results}
All training agents are implemented using the actor-critic algorithm with Proximal Policy Optimization (PPO) \cite{Schulman_ppo}.
The network architecture is detailed in Appendix \ref{append:network_parameter}.

\subsection{Curriculum Learning Stage}
As described in Section \ref{sec:approach_curriculum}, in this stage, the agent will learn essential skills by training with three different rule-based agents in three phases. The adaptive exploration reward by performanc is also applied in this stage.

In the beginning, the agent does not know how to eliminate the opponents, so the annealing factor $\alpha$ in Equation \ref{equal:adaptive_reward} is equal to 0.
In this way, the agent prefers to do actions to explore the area as well as pick up items while it is not punished by the negative reward of accidental suicide.

After about 5 million timesteps, the agent is able to locate and eliminate \texttt{static\_agent}, reaching the win rate of 55\%.
Then, the second phase with \texttt{simple\_moving\_agent} of curriculum learning is activated.
Due to the increase in difficulty in the second phase, the performance of the agent quickly drops.
Following the decrease in performance is the increase in the annealing factor, which forces the agent to pick up items and place more bombs to find better strategies to eliminate the enemies.

Finally, in the third phase,  the \texttt{simple\_bombs\_agent} starts to place bombs randomly, making it harder for the training agent to approach the opponent. 
As a result, the training agent learned defensive strategies to effectively dodge the bomb's explosion while simultaneously threatening and eliminating opponents.

\begin{figure}[!ht]
\includegraphics[width=0.6\textwidth]{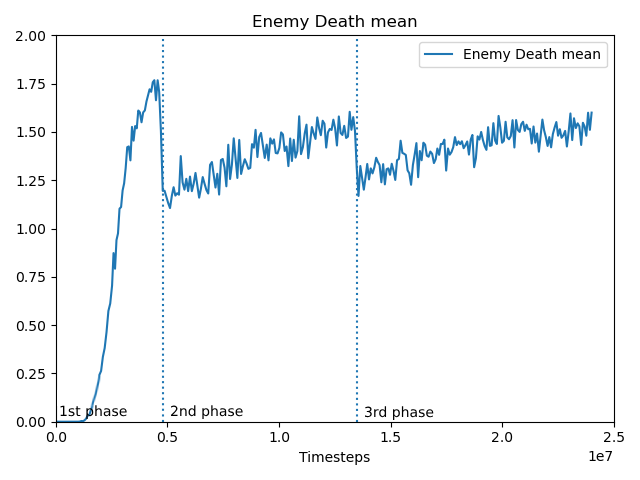}
\centering

\caption{Average of Enemy Deaths.}
\label{fig:result_1}
\end{figure}

\begin{figure}[!ht]
\includegraphics[width=0.6\textwidth]{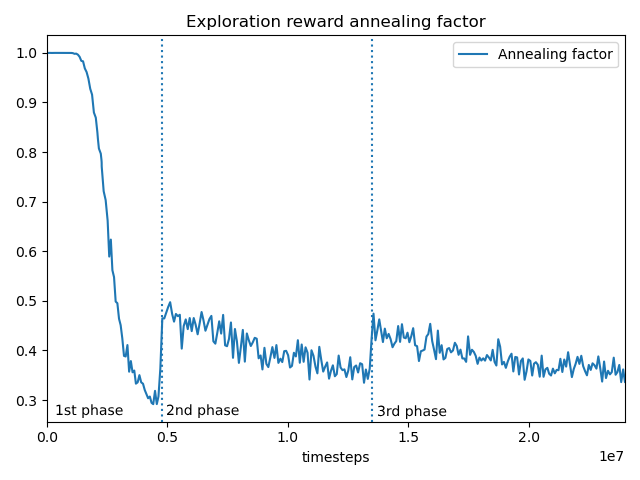}
\centering

\caption{Exploration reward annealing during the training process.}
\label{fig:result_2}
\end{figure}

The two Figure \ref {fig:result_1} and Figure \ref {fig:result_2} illustrate the contrast between the annealing factor and the performance of the training agent. 

\begin{figure}[!ht]
\includegraphics[width=0.6\textwidth]{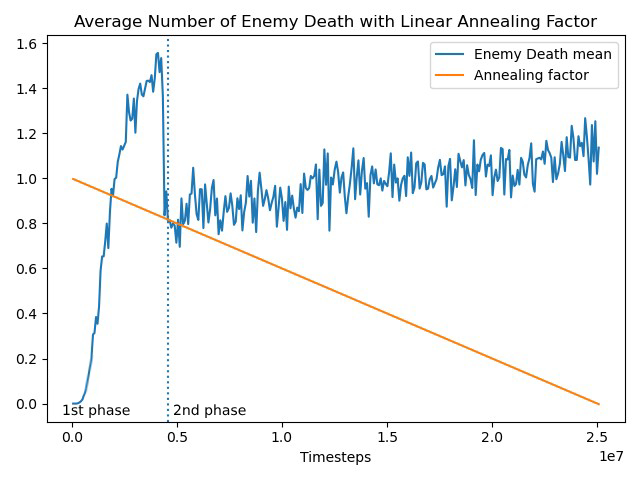}
\centering
\caption{Average number of Enemy deaths using linear annealing factor.}
\label{fig:result_3}
\end{figure}

In comparison with the performance annealing factor, we train another model with a linear annealing factor to show the difference between the two methods.
Figure \ref{fig:result_3} illustrates the average number of enemy deaths together with the linear annealing factor.
After around 5 million timesteps, the training agent can also finish the first phase and activate the second phase, the same as in the performance of the adaptive exploration reward method.
In the next phase, the performance of the training drops immediately.
The problem is that the annealing factor cannot adjust to encourage the training agent to keep exploring.
Ultimately, the training agent takes a long time to play against the second rule-based agent and cannot pass the second phase.
Note that adjusting the value of the linear annealing factor can lead to improved results.
However, it will introduce another problem of manually fine-tuning the hyperparameters.

After the curriculum learning stage, the training learned useful behaviors such as exploding wood walls, picking up revealed items, dodging, hiding from the explosion, and finding enemies.
Also, the dense exploration reward is annealed to zero.
The reward function in the self-play stage now contains only sparse game rewards, allowing the training agent to self-develop their strategy aligned with the main game objective.

\subsection{Self-play Stage}
As described in Section \ref{sec:approach_self_play}, the self-play stage utilizes a population-based training system with eight agents: three rule-based agents (\texttt{static\_agent}, \texttt{simple\_move\_agent}, \texttt{ simple\_bombs\_agent}), and five training agents (initialized from the trained agent in the curriculum stage).
We evaluate the agent using the Elo rating.
Initially, every agent is given 1000 Elo rating points and continuously updates after every match.
This Elo rating point is also used to obtain matchmaking probability between two agents, using the mechanism detailed in Subsection \ref{sec:matchmaking}, which ensures the selection of suitable opponents for facilitating progressive learning and avoids getting stuck.

\begin{figure}[!ht]
\includegraphics[width=0.9\textwidth]{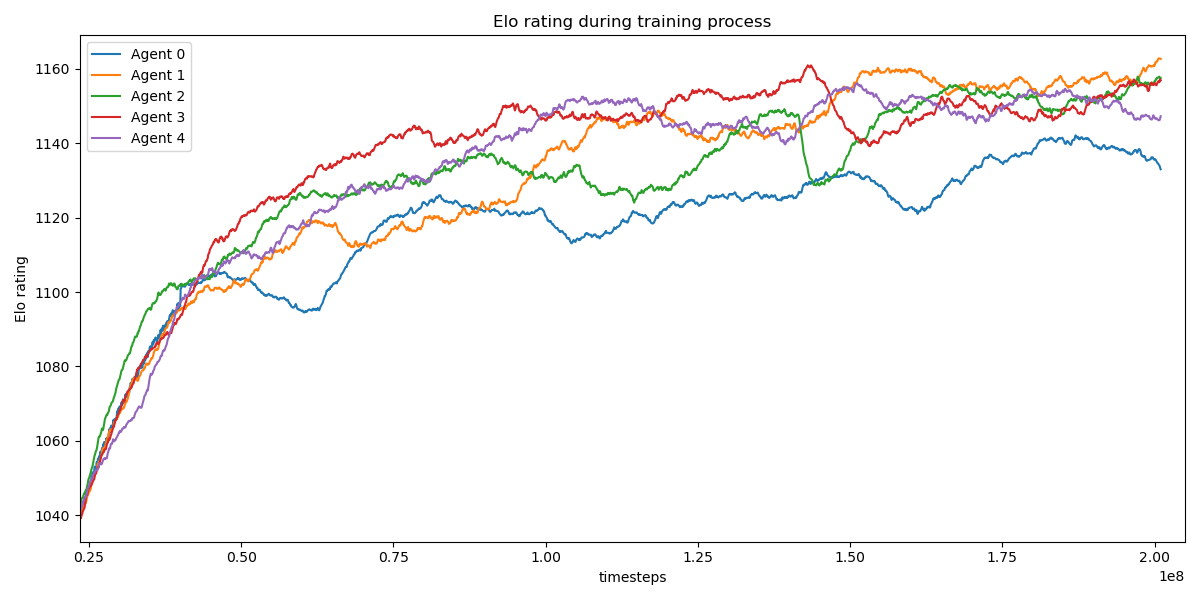}
\centering

\caption{Elo rating of five training agents during the self-play training stage.}
\label{fig:training_chart}
\end{figure}

Figure \ref{fig:training_chart} demonstrates the Elo rating of our training agent during the self-play training stage.
After the curriculum stage, our agents start with 1040 Elo rating points.
The Elo rating keeps increasing gradually to 1160 after 200 million training timesteps.

\begin{figure}[!ht]
\includegraphics[width=0.99\textwidth]{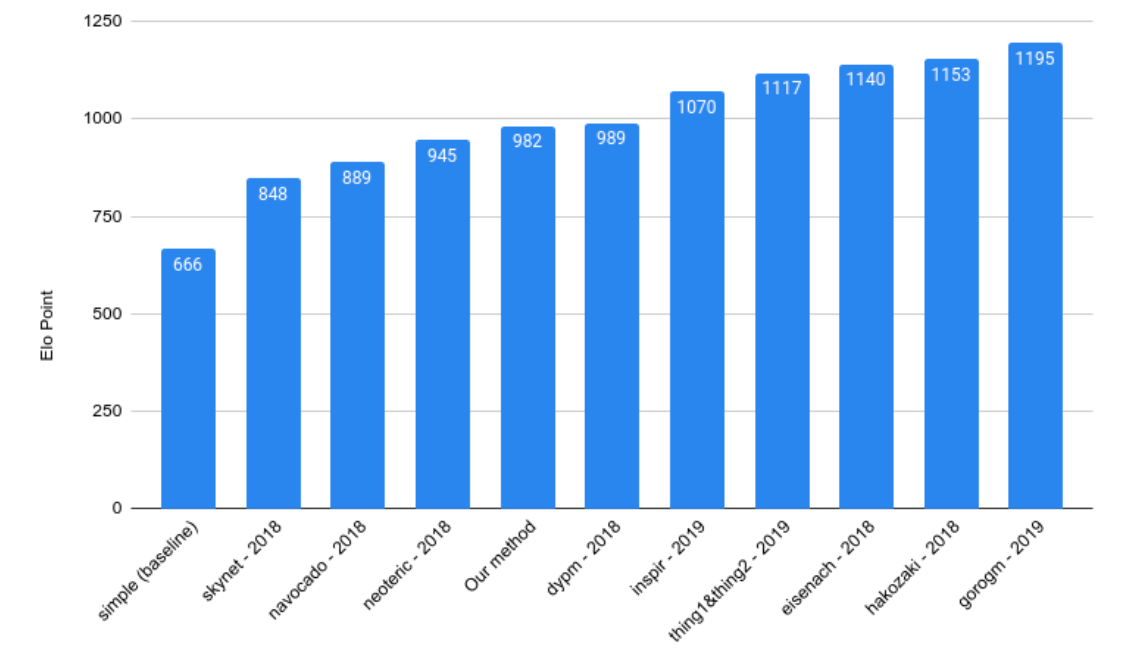}
\centering

\caption{Evaluating Elo rating of the agents with 100 round-robins matches.}
\label{fig:elo_chart}
\end{figure}

After finishing self-play stage training, we arranged 100 round-robin matches to evaluate all methods, including a baseline agent developed by Pommerman, nine agents from the 2018 and 2019 competitions, and our agent.
As shown in Figure \ref{fig:elo_chart}, our method has 982 Elo rating points, which is greater than the top two learning agents in 2018 and a robust rule-based agent (Neoteric in 2019).
Besides, our Elo rating almost equals to \textit{dypm}, a tree search-based agent.

\begin{table}[!ht]
\centering
\begin{tabular}{|l|l|l|l|l|}
\hline
            & simple (baseline)  & skynet  & navocado & neoteric \\ \hline
Our win rate & 98.85\% & 96.23\% & 88.33\%  & 87.69\%  \\ \hline
\end{tabular}

\caption{Win rate of our method against other learning agents}
\label{tab:winrate_1}
\end{table}

\begin{table}[!ht]
\centering
\begin{tabular}{|l|l|l|l|l|l|l|}
\hline
            & dypm    & eisenach & hakozaki & gorogm & inspir & thing1 \\ \hline
Our win rate & 30.00\% & 8.89\%   & 5.17\%   & 4.35\% & 4.00\% & 1.75\% \\ \hline
\end{tabular}

\caption{Win rate of our method against tree-search-based agents and communicated agents.}
\label{tab:winrate_2}
\end{table}

We also compare the win rate of our agent against other agents.
As shown in Table \ref{tab:winrate_1}, our agent has a win rate of 98.85\% against the baseline of Pommerman, which is a heuristic agent using the Dijkstra algorithm to find the opponents and place bombs next to enemies.
Also, our agent outperforms the top learning agents in 2018, Skynet955 and Navocado, with 96.23\% and 88.33\% win rate, respectively.
Furthermore, we defeated Neoteric - a ranked 4th agent in 2019, using robust rule-based strategies.
Our agent is defeated when facing tree-search-based agents or methods that have communication in their team (shown in Table \ref{tab:winrate_2}), which is out of scope in this study.

\section{Conclusions}
This research introduces a training system for multi-agent learning in Pommerman with two stages: curriculum learning and population-based self-play.
After the curriculum learning stage, the agent learned essential skills such as placing bombs to explode wooden walls and picking up revealed items.
After the self-play stage, the agents' strength is improved, and they autonomously learn a more effective and secure strategy, which includes strategic behaviors such as kicking bombs toward enemies or trapping enemies with bombs.
We also address two challenges when deploying a multi-agent self-play training system for competitive games: sparse reward and suitable matchmaking.
Specifically, we propose an adaptive annealing factor based on agents’ performance to dynamically adjust the dense exploration reward, gradually prioritizing the game reward.
Furthermore, we implement a matchmaking mechanism utilizing the Elo rating system to pair agents effectively, ensuring incremental learning during the self-play training.
Finally, without communication, our trained agent is able to defeat the top learning agent and the top four rule-based agents in the 2019 competition.

\clearpage


\bibliography{example}  

\clearpage
\newpage

\appendix
\section{The Pommerman}
\subsection{Game Description}
\begin{figure}[h!]
\includegraphics[width=0.5\textwidth]{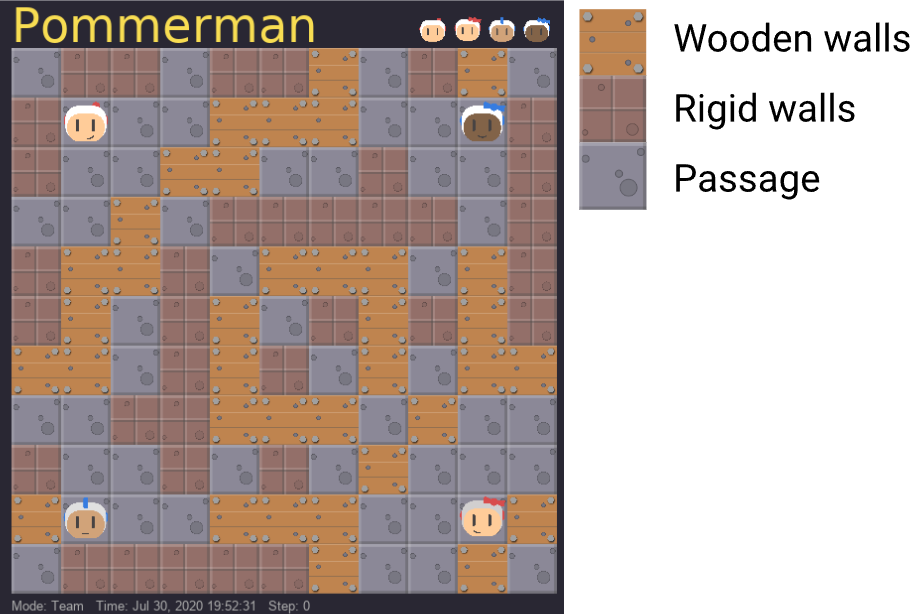}
\centering
\captionsetup{justification=centering}
\caption{Initial game board, where four agents are located at four corners.}
\label{fig:board_1}
\end{figure}

In Pommerman, there are four agents, and each agent is placed in four corners.
The board size is 11 x 11. It is symmetric along the main diagonal.
The board is scattered with three types of cells: wooden walls, rigid walls, and passages (see Figure \ref{fig:board_1}).

\begin{figure}[h!]
\includegraphics[width=0.7\textwidth]{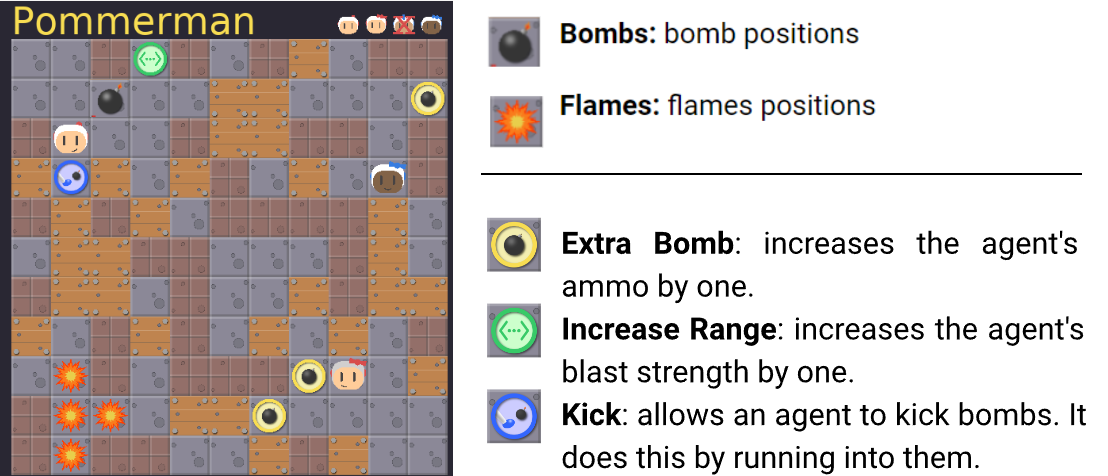}
\centering
\captionsetup{justification=centering}
\caption{Game board with bombs, flames, pick-up items.}
\label{fig:board_2}
\end{figure}

In general, each agent can place a bomb that will explode after 10 timesteps later.
Flames are created after a bomb explodes, which is in a cross shape with a size of 1 cell around the exploded bomb.
The flames eliminate any agents, wooden walls,
and items if they hit the objects and explode other bombs on collision.
Furthermore, agents cannot go through bombs and obstacles. Wooden walls, after destruction, may reveal items for agents to pick up.
There are three types of pick-up items, which are the extra bomb (increases the agent's ammunition by one), increased range (increases the size of bomb explosion of an agent by one), and kick-ability (allows the agent to kick bombs) (see Figure \ref{fig:board_2}).

\subsection{Game Modes}
There are two unique modes in Pommerman: Free-For-All mode (FFA) and team mode.

In the first mode, all agents in a battle fight against each other, and the last survival wins the game.

Team mode is a 2-versus-2 setting where one team needs to eliminate the other team.
Also, the board is partially observable because the view of the agent is limited in a 9x9 grid around the agent.
The allied agents are allowed to communicate with each other by sending a list of two numbers in the range of 1 to 8. 
If at least one member of each team survives at the end of the battle, the result will be a tie.

\section{State Representation}
Pommerman internal states are represented into 2D features and scalar features.
The game is partial observation, with a limit observation size of 9x9 around the player position, so the 2D features are spatially related features size 9x9 with the center being the player position.

Scalar features present some information about the game, like ammunition, bomb blast straight, and equipped item ability.

\begin{figure}[h!]
\includegraphics[width=0.99\textwidth]{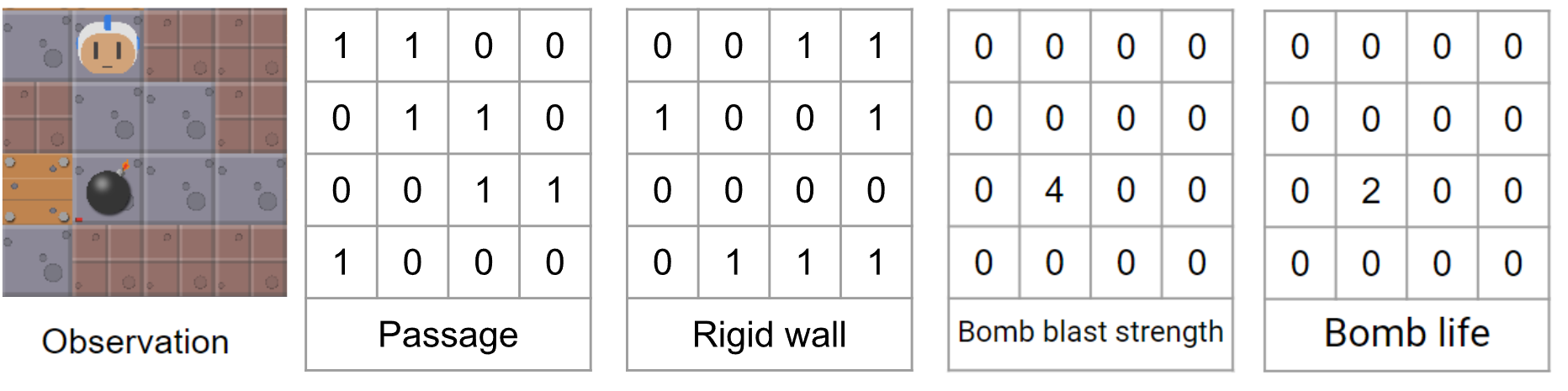}
\centering
\captionsetup{justification=centering}
\caption{Example of 2D features.}
\label{fig:2D_features_example}
\end{figure}

Figure \ref{fig:2D_features_example} is an example of a map of a 4x4 grid. The figure represents passages and rigid walls in binary features for any available cell. Besides, the bomb blast strength is 4, and the bomb life is 2, which illustrates the strength and
time-to-explode of the bomb on the board.
In particular, a state in every step is represented by a 9x9 grid around the agent
which is the same as the limited vision of the agent.
The cells outside of the board are filled with rigid walls.
The state represented in this way is more compact and helps an agent focus on surrounding objects and speed
up the training process.

Table \ref{tab:2D_features} and Table \ref{tab:scalar_feature} are 2D features and scalar features used in our research.

\begin{table}[h!]
\centering
\begin{tabular}{|l|l|}
 \hline
\textbf{Features}     & \textbf{Describe}                              \\ \hline
Passage               & Indicate the passege path                      \\ \hline
Rigid walls           & Indicate the position of rigid walls           \\ \hline
Wooden walls          & Indicate the position of wooden walls          \\ \hline
Bombs                 & Indicate the position of bombs are placed      \\ \hline
Flames                & Indicate the position of flames                \\ \hline
Enemy position        & Indicate the position of enemy                 \\ \hline
Current position      & Indicate the position of current player        \\ \hline
Teammate position     & Indicate the position of teammate              \\ \hline
Item – Increase range & Indicate the position of Item - Increase range \\ \hline
Item – Extra bomb     & Indicate the position of Item - Extra bomb     \\ \hline
Item – Kick           & Indicate the position of Item-Kick             \\ \hline
Bomb blast strength   & The blast strength of the placed bombs.                                               \\ \hline
Bomb life             &  The remaining time before the bombs explode.                                             \\ \hline
\end{tabular}
\captionsetup{justification=centering}
\caption{2D features (size 9x9) of Pommerman}
\label{tab:2D_features}
\end{table}

\begin{table}[h!]
\centering
\begin{tabular}{|l|c|l|}
\hline
\textbf{Features}   & \textbf{Shape} & \multicolumn{1}{c|}{\textbf{Describe}} \\ \hline
Position            & 2              &   2D coordinate of the current player                                     \\ \hline
Ammunition          & 1              &     The available number of bombs to place.                                   \\ \hline
Bomb blast strength & 1              &      The blast strength of the bomb when exploding.                                  \\ \hline
Kick ability        & 1              &      Indicate that the player has kick-ability or not                                  \\ \hline
Teammate alive      & 1              &   Indicate that teammate is alive or not                                     \\ \hline
Two opponents         & 1              &   Indicate the number of opponents.                                     \\ \hline
\end{tabular}
\captionsetup{justification=centering}
\caption{ Scalar features of Pommerman}
\label{tab:scalar_feature}
\end{table}

\section{Reward System}
\label{sec:reward_system}
The main objective of Pommerman is to eliminate the opponents.
However, players are initially located in the corners of the map and are blocked by wooden walls and stones.
Thus, the gameplay can be divided into two subtasks: exploring the map to navigate to opponents and engaging in combat to eliminate them.

Corresponding to the two subtasks, we define two reward types: exploitation reward $e_t$ and game reward $R$.
The exploration reward $e_t$ is received at timestep $t$ to encourage the agent to place bombs to explore the map and uncover the items hidden behind the wooden walls.
The game reward $R$ is received at the end of the game based on the game results (win, tie or loss), motivating the agent to engage in combat and eliminate opponents.
The reward system for training is defined as follows:
\begin{table}[h!]
\begin{tabular}{|c|l|l|}
\hline
\multirow{2}{*}{Exploration reward $e_t$} & Pick up items (blast strength, ammo, kick) & +0.1                 \\ \cline{2-3} 
                                    & Placing bomb                               & +0.005               \\ \hline
\multirow{3}{*}{Game reward $R$}        & At the end of episode                      & +1/death enemy \\ \cline{2-3} 
                                    & Death                                      & -1                   \\ \cline{2-3} 
                                    & Tie                                        & -1                   \\ \hline
\end{tabular}
\captionsetup{justification=centering}
\caption{Reward system for training}
\label{tab:reward_system}
\end{table}

\section{Network Architecture and Training Parameters}
\label{append:network_parameter}

All training agents are implemented using the actor-critic algorithm with Proximal Policy Optimization (PPO).
Our network architecture is conducted from 4 convolutional layers, which have kernel sizes of 3x3, stride 1, padding 0, and are activated by a ReLU function.
At the end of the final convolutional layer, the output is flattened into 256 units and
concatenated with scalar features.
After that, the extracted features were put into the Long Short Term Memory (LSTM) \cite{Hochreiter_lstm} layer to remember the information in a sequence of 10 steps.
The output of the LSTM layer is a fully connected layer of 128 units, which splits into two heads: value head and policy head.
A value head estimates the value of the current state, and a policy head outputs the probabilities of 6 actions (stay, up, down, left, right, bomb) of the agent.

\begin{figure}[h!]
\includegraphics[width=0.99\textwidth]{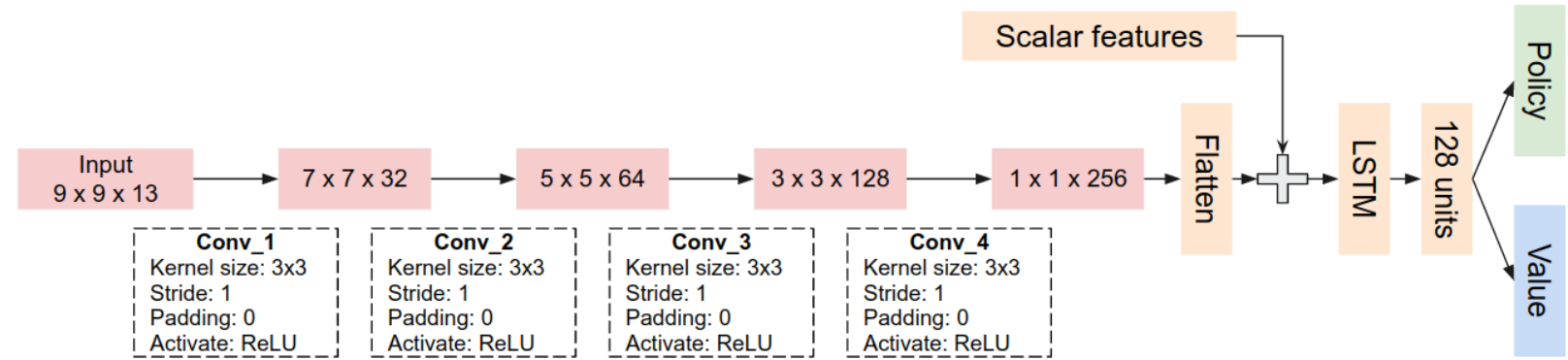}
\centering
\captionsetup{justification=centering}
\caption{Network structure.}
\label{fig:network_structure}
\end{figure}

Figure \ref {fig:network_structure} visualizes our network structure used in this research.


\end{document}